\documentclass[%
 amsmath,amssymb,
 aps,
prl,
twocolumn,
]{revtex4-2}
\usepackage{floatrow}
\usepackage{graphicx}
\usepackage{dcolumn}
\usepackage{bm}
\usepackage[labelfont=bf,labelformat=parens]{subfig}
\usepackage{gensymb}
\usepackage{xcolor}
\usepackage{hyperref}

\begin{document}

\preprint{APS/123-QED}

\title{Absence of high pressure ground state re-entrant ferroelectricity in PbTiO$_3$} 

\author{R. E. Cohen}
\email{rcohen@carnegiescience.edu}
\author{Yangzheng Lin}
\affiliation{Extreme Materials Initiative, Earth \& Planets Laboratory, Carnegie Institution for Science, Washington DC 20015, USA}
\author{Muhtar Ahart}%
\email{maihaiti@uic.edu}
\affiliation{Department of Physics, University of Illinois Chicago, Chicago IL 60607, USA}
\author{Russell J. Hemley}
\email{rhemley@uic.edu}
\affiliation{Departments of Physics, Chemistry, and Earth and Environmental Sciences,        
University of Illinois Chicago, Chicago IL 60607, USA}

\date{\today}

\begin{abstract}
We study ferroelectricity in the classic perovskite ferroelectric PbTiO$_3$ to high pressures with density functional theory (DFT) and experimental diamond-anvil techniques. We use second harmonic generation (SHG) spectroscopy to detect lack of inversion symmetry. Consistent with early understanding and experiments, we find that ferroelectricity disappears at moderate pressures. However, DFT computations show that the disappearance arises from the overtaking of zone boundary instabilities, and not from the squeezing out of the off-centering ferroelectric displacements with pressure, as previously thought. Moreover, at high pressures the distorted perovskite phases are metastable with respect to a new dense centrosymmetric post-perovskite phase with P$2_1/m$ symmetry and 8-coordinated Ti, which becomes stable at about 70 GPa. 
\end{abstract}

\maketitle


\begin{figure}
\includegraphics[scale=0.1]{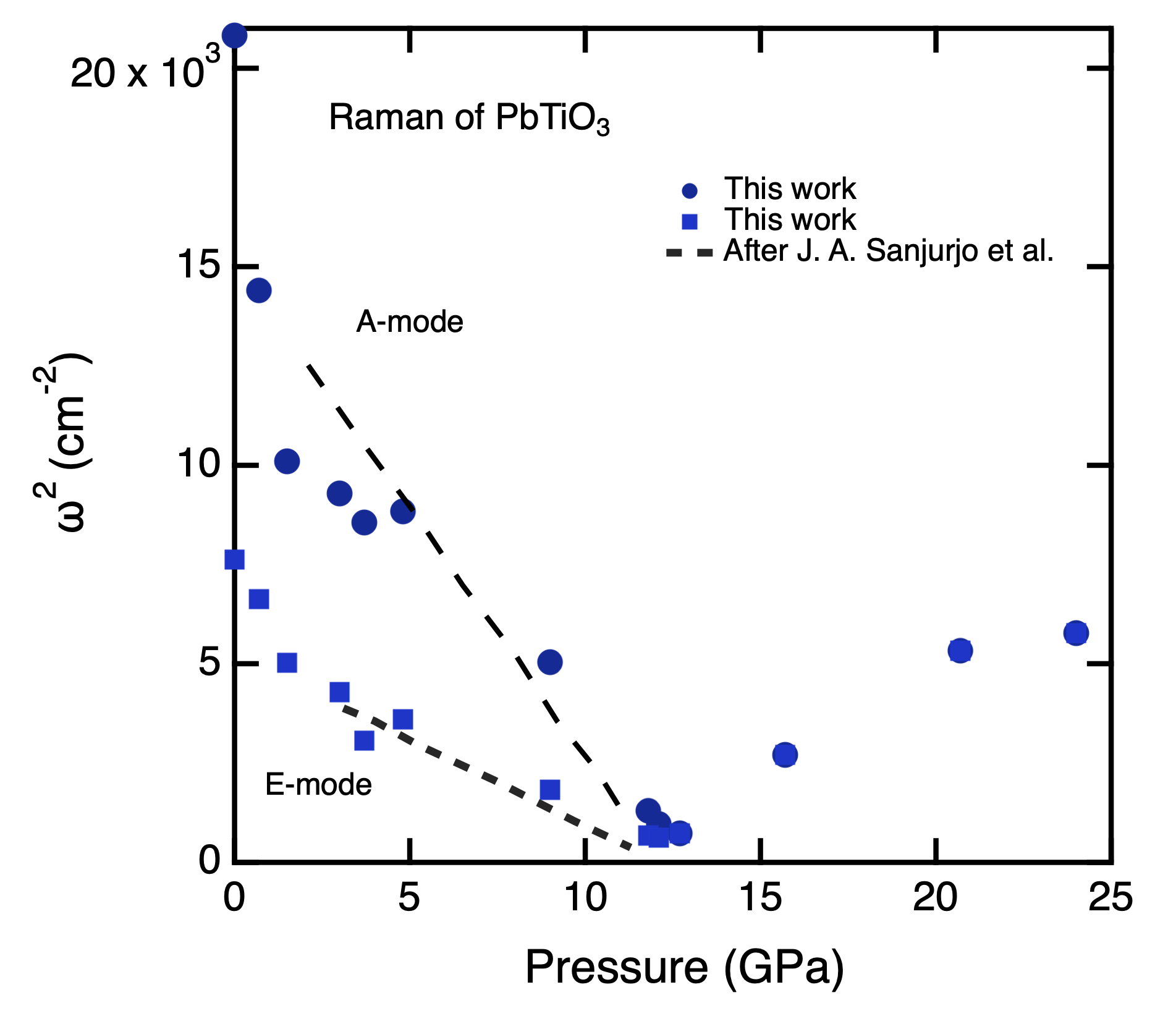}
\caption{\label{Raman}Raman and energetics of low-pressure PbTiO$_3$. Experimental room temperature low frequency Raman. Blue symbols are our measured data at room temperature (300 K). Dashed curves are experiments from Ref.~\cite{Sanjurjo1983}, showing the mode softening with pressure leading to what was believed to be cubic PbTiO$_3$, with no Raman spectrum, at the moderate pressure of 12 GPa. We also observe a weaker mode which hardens with pressure.}
\end{figure}

Ferroelectricity was long thought to be a low-pressure phenomenon; for example non-existent above 12 GPa in PbTiO$_3$ \cite{Sanjurjo1983}
. This also fit with the longstanding belief that crystal structures become simpler and more symmetric with increasing pressure \cite{Hemley2010}. This concept is intuitive, since with increasing pressure atomic density must increase, and hard spheres tend towards close packing with increasing density.  Although numerous high-pressure studies conducted over the past two decades contradict this conjecture, it still seemed clear that ferroelectricity, which requires a departure from atomic close-packing, would disappear with increasing pressure, especially when thought of in terms of Slater's rattling ion model \cite{Slater1950}, still popular in textbooks, and its underlying view that there is simply less room for ions to rattle as density increases. 

In contrast to the above, we have long known that covalency or hybridization is important in allowing ions to off-center \cite{Cohen1992}, so one could imagine pressure-induced covalency leading to regimes of high-pressure ferroelectricity. This effect was predicted by density function theory (DFT) calculations for PbTiO$_3$ \cite{Kornev2005}, with experimental confirmation claimed \cite{Janolin2008} and interpreted as arising from hybridization of Ti $d$ states with O 2$s$ states \cite{Kornev2007}. Other DFT studies examined phase stability under pressure \cite{Ganesh2009} and showed that the results are sensitive to small energy differences arising from the choice of pseudopotential and/or exchange-correlation potential, but are generally consistent with these findings \cite{Kornev2005,Kornev2007}. High-pressure Raman and x-ray measurements measured at room temperature \cite{Janolin2008} were interpreted as showing a sequence of transitions from tetragonal $P4mm$ to possibly cubic perovskite at 10-14 GPa, followed by tetragonal $I4/mcm$ at 18-20 GPa to polar $I4cm$ at 37-50 GPa. DFT computations and some preliminary experiments were presented for PbZrO$_3$ under pressure \cite{Prosandeev2014}. Although the R3c structure in that paper is relevant to this work on PbTiO$_3$, the other phases are never stable in PbTiO$_3$.

Here we call into question the idea of reentrant ferroelectricity in perovskites. In this paper we revisit the theoretical calculations of PbTiO$_3$ and present SHG experiments to reveal the complexity of the high pressure phases, including a new centrosymmetric structure and the absence of high pressure, ground state ferroelectricity.

Although room temperature experiments showed an apparent simple soft mode transition to cubic at 12 GPa \cite{Sanjurjo1983}, low-temperature (10 K) measurements\cite{Ahart2008} revealed a transition region at 10-14 GPa, characteristic of a morphotropic phase boundary, from tetragonal to monoclinic and then rhombohedral symmetry, consistent with DFT predictions \cite{Wu2005}. 

Room-temperature x-ray and Raman measurements indicated that PbTiO$_3$ is not cubic at higher pressures \cite{Janolin2008}, but the results could not show whether or not the high-pressure phase was polar, because conventional x-ray diffraction cannot unambiguously determine the presence of inversion symmetry. Furthermore, in powder x-ray diffraction only changes in lattice parameters are readily discerned, and typically simulations are needed to test for changes in intensities from small atomic displacements, especially considering possible preferred orientation and pressure gradients. 

To test for loss of inversion symmetry, which is necessary for ferroelectricity, we thus conducted a series of second harmonic generation (SHG) spectroscopy measurements. SHG is a sensitive test for the presence or absence of inversion symmetry. Any ferroelectric below T$_c$ should give a significant SHG signal.

\begin{figure}
\includegraphics[scale=0.2]{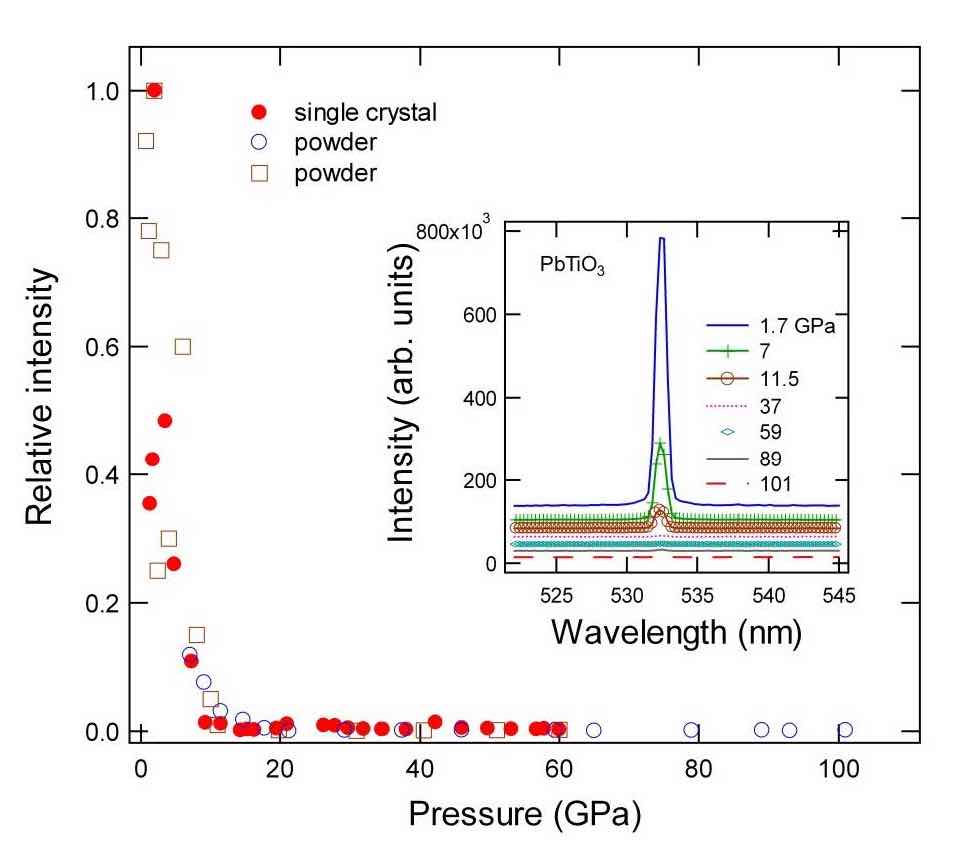}
\caption{\label{SHG}SHG signal for PbTiO$_3$ versus pressure. Powder data at room temperature to 100 GPa and single crystal data and powder data at 10 K to 60 GPa are shown. Above 20 GPa no SHG signal is observed, indicating the sample is not polar, consistent with theory. A Ne pressure medium  maintained quasihydrostatic conditions.}
\end{figure}

We studied PbTiO$_3$ Single crystals grown by a high-temperature flux technique. X-ray diffraction confirmed that samples had $P$4$mm$ symmetry at ambient conditions. Angle dispersive x-ray diffraction experiments were carried out at beamline 16-IDB of HPCAT, Advanced Photon Source, Argonne National Laboratory. For the SHG measurements we used a near IR (1064 $\mu$m), 8-20 ns pulsed laser with a 1-20 kHz repetition rate.  A spectrograph equipped with a charge-coupled-device (CCD) detector synchronized with the laser was used the detect the signal. The acquisition time was typically about 2 s. Single crystal and powder samples were measured in three-run experiments. See Supplement for more details on the experiments \cite{SM}. At room temperature, we found that the SHG signal disappears above 10-12 GPa (Fig.~\ref{SHG}), consistent with the earlier Raman measurements \cite{Sanjurjo1983}. At 10 K, the signal decreased approximately linearly to 15 GPa, and then decreases in stages to zero at 25 GPa. The results are consistent with a transition to a centrosymmetric phase.

\begin{figure}
\includegraphics[scale=0.07]{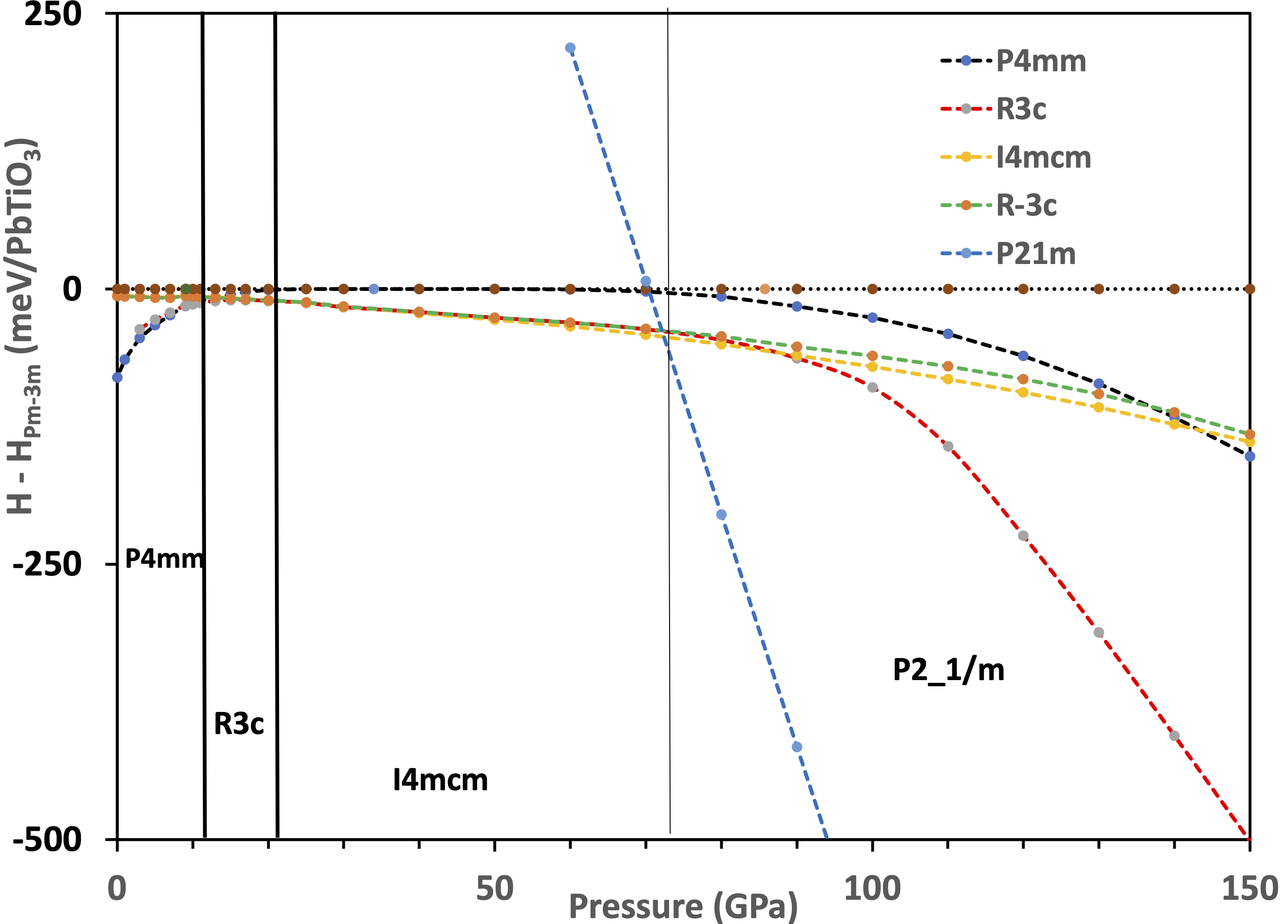}
\caption{\label{H}Computed enthalpy differences of PbTiO$_3$ phases relative to cubic perovskite versus pressure. The ground-state stability fields of the most stable phases are labeled.}
\end{figure}

\begin{figure}
\includegraphics[scale=0.06]{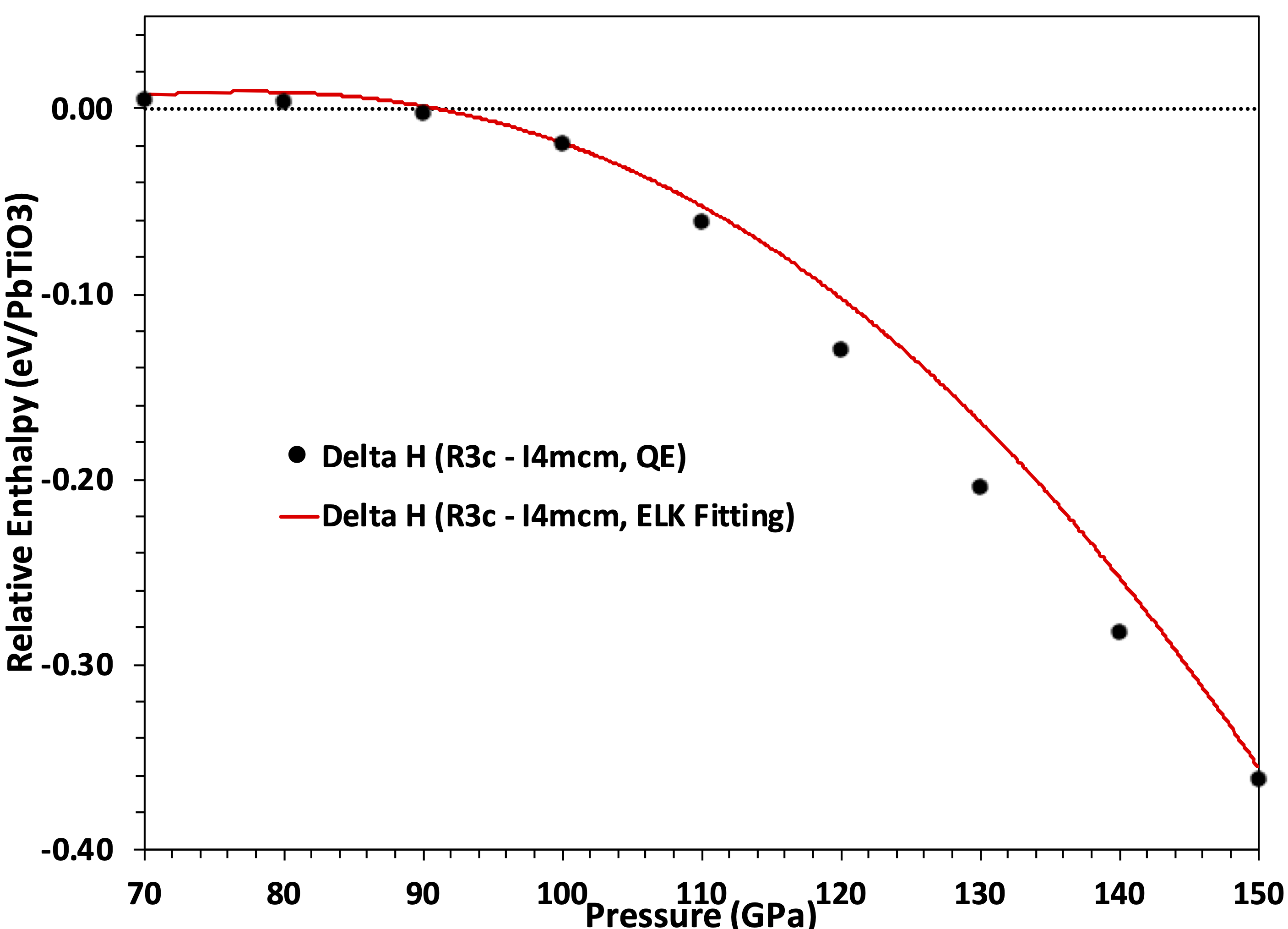}
\caption{\label{compare}Comparison of enthalpy differences from $I4/mcm$ to $R3c$ for PbTiO$_3$ computed with pseudopotentials and plane waves (points) and all-electron full-potential LAPW (solid line from an equation of state fit for same structure parameters as points).}
\end{figure}

To understand better the experimental results, and the apparent discrepancy with the previous DFT computations, we performed computations for PbTiO$_3$ under compression using both pseudopotentials using \textsc{Quantum Espresso} \cite{QE} and all-electron LAPW or APW+LO using the \textsc{ELK} code \cite{ELK}. We generated norm conserving ONCV pseudopotentials \cite{Hamann2013} using the Wu-Cohen exchange correlation potential \cite{Wu2006}. Tests for the optimized structures showed the same energy difference from all-electron computations with ELK (Fig.~\ref{compare}), but optimizations were more convenient with accurate analytic stresses in the pseudopotential computations. The enthalpies of the relaxed structures show a sequence of transitions from $P4mm \rightarrow R3c \rightarrow  R\overline{3}c \rightarrow I4/mcm \rightarrow P2_1/m$ (Fig.~\ref{H}). Above 15-20 GPa the structures are all centrosymmetric and thus there is no high-pressure ferroelectricity. The detailed behavior between 10 and 20 GPa is complicated, involving very small enthalpy differences (<1 meV/atom) between polar 10 atom/cell $R3c$ with a very small polarization, centrosymmetric $R\overline{3}c$, and monoclinic polar 5 atom cells Cm and Pm, and we do not try to untangle that here as the phase diagram here will depend on temperature and anharmonic contributions to the free energy. With very high cut-offs and k-point sets and the WC exchange-correlation functional, $R3c$ is the lowest enthalpy phase at static zero temperature, so we show that here.

\begin{figure}
      \centering
\subfloat[$Pm3m$]{\includegraphics[width=0.3\linewidth]{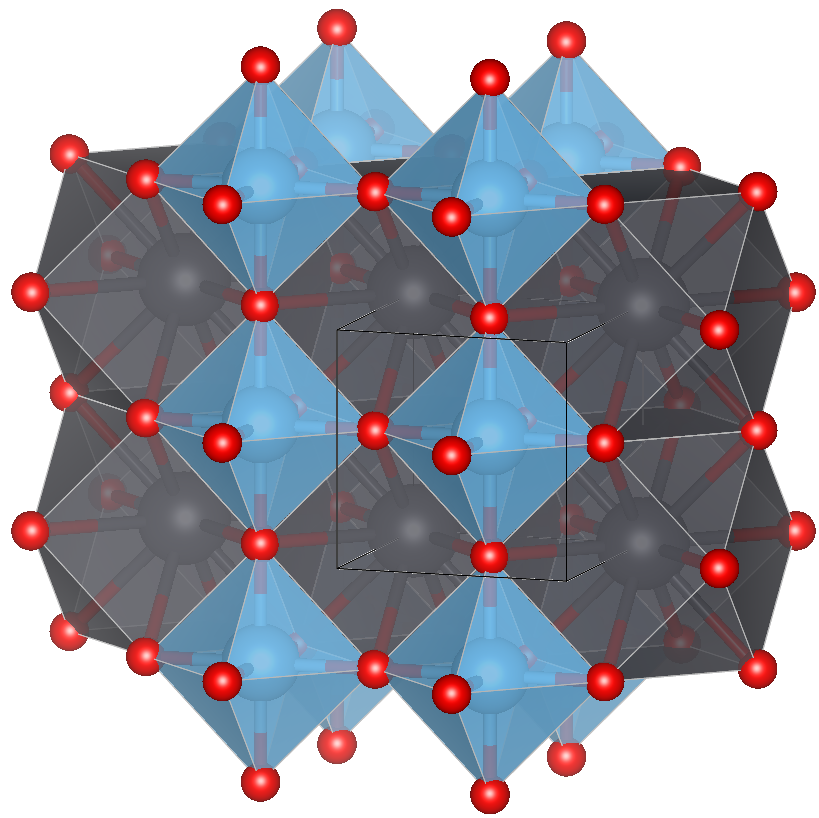}\label{fig:Pm3m}}
\subfloat[$P4mm$]{\includegraphics[width=0.3\linewidth]{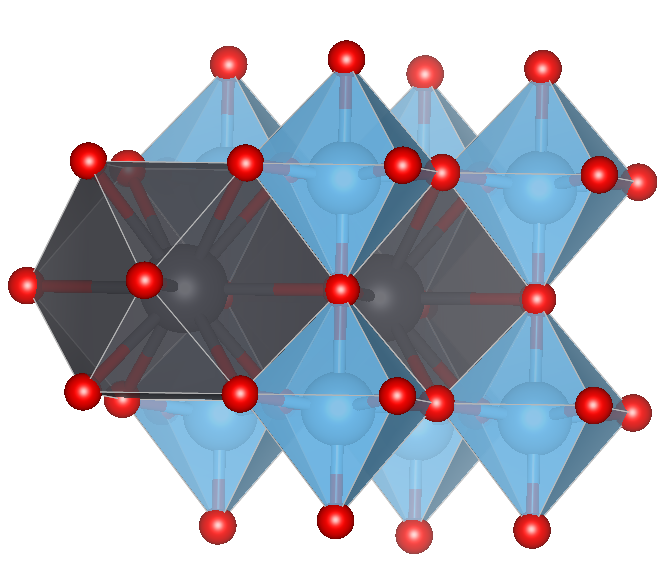}\label{fig:P4mm}}
\subfloat[$R3c$]{\includegraphics[width=0.3\linewidth]{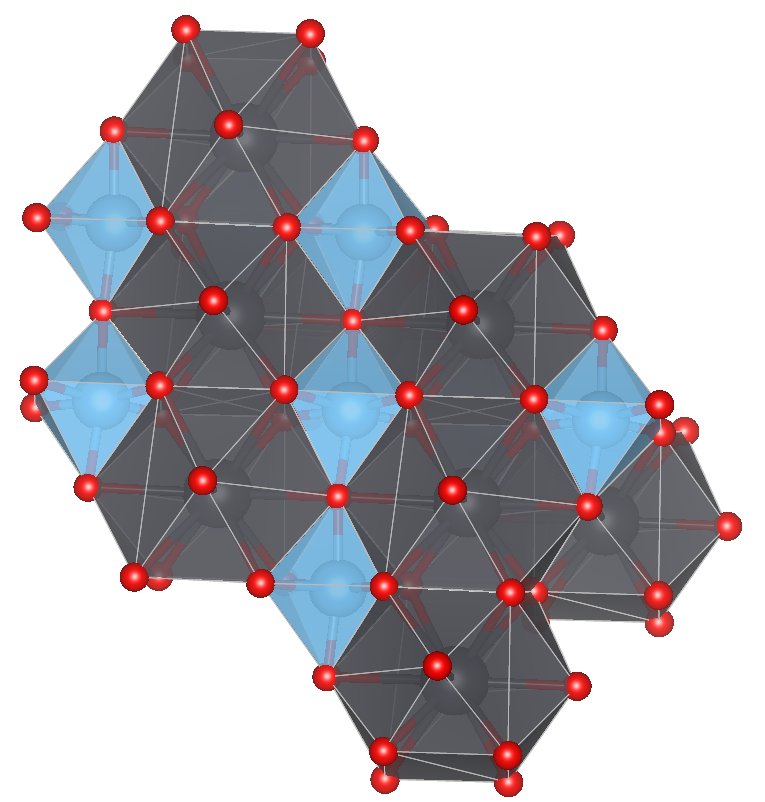}\label{fig:R3c}}
\\
\subfloat[$R\overline{3}c$]{\includegraphics[width=0.3\linewidth]{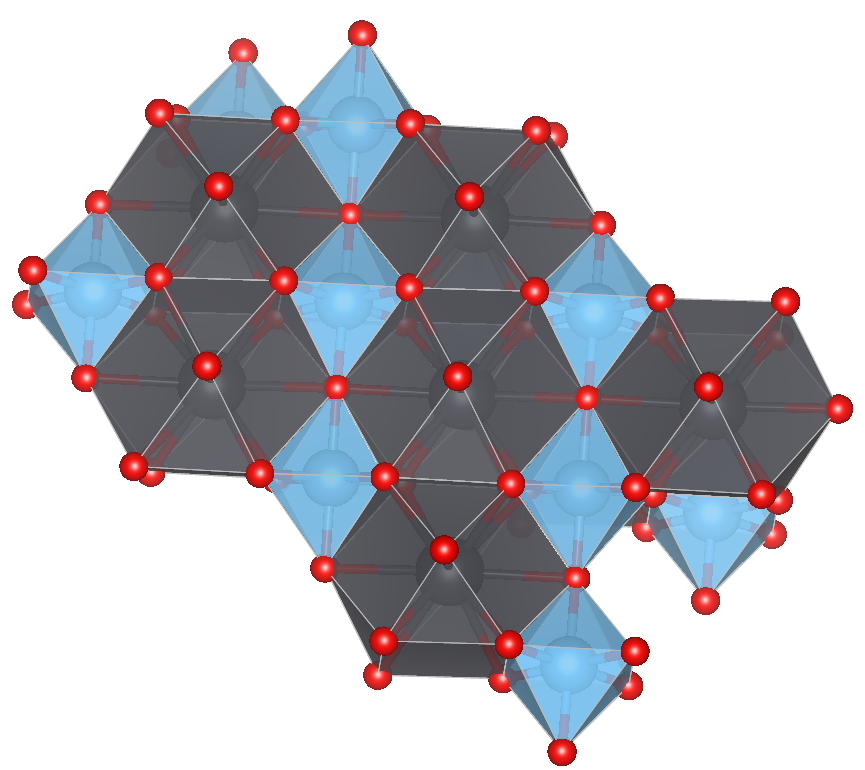}\label{fig:R3-c}}
\subfloat[$I4/mcm$]{\includegraphics[width=0.3\linewidth]{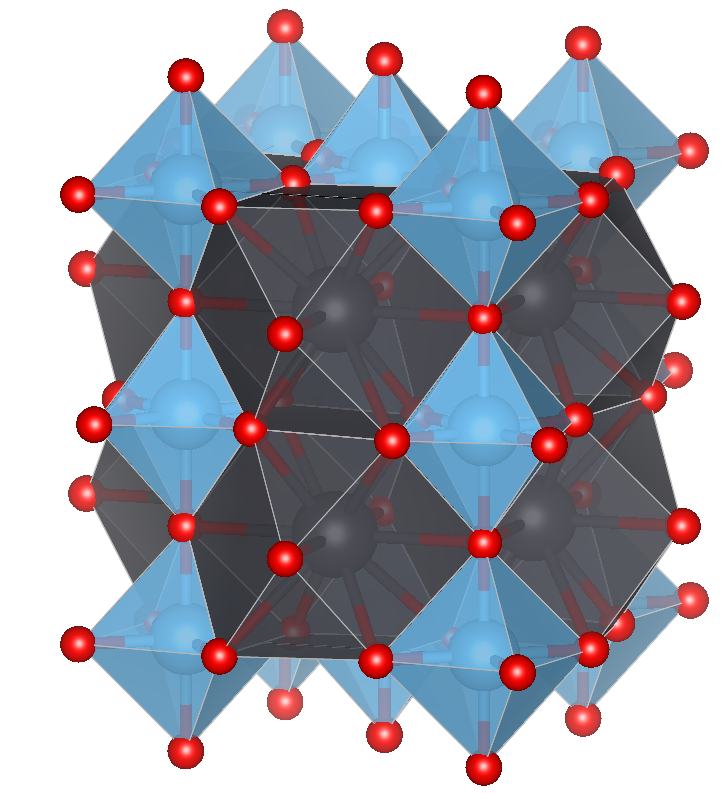}\label{fig:I4mcm}}
\subfloat[$P2_1/m$]{\includegraphics[width=0.3\linewidth]{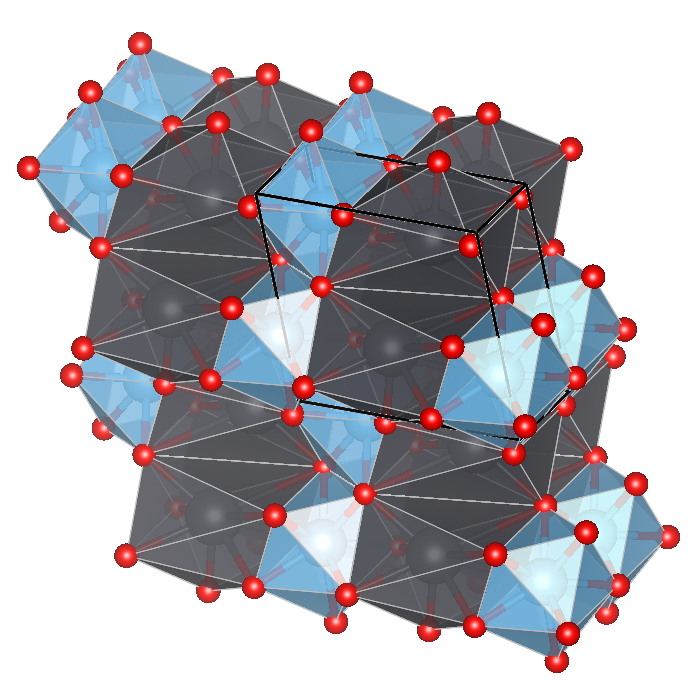}\label{fig:P21m}}
\caption{Crystal structures for PbTiO$_3$, all shown for structures relaxed at 80 GPa.}
	\label{fig:PTO_structures}
\end{figure}

Considering firstly only perovskite derivative structures, we found a polar $R3c$ phase at about 86 GPa, in contradiction to our highest pressure experiments, which did not show any SHG signal. To understand the origin of this discrepancy, we undertook a structure search using \textsc{XtalOpt}\cite{Lonie2011,Falls2020} and \textsc{Quantum Espresso} at 150 GPa, and discovered a new phase with 10 atoms per unit cell and $P2_1/m$ symmetry (Figs.~\ref{H},\ref{fig:P21m}).  Optimizing this structure versus pressure and comparing with the other structures, we found the new phase to become the ground state at 72 GPa (Fig.~\ref{H}). At 80 GPa a=5.52 \AA, b=2.71 \AA, and c=5.92 \AA, and $\beta$=98.614 \degree. The atoms are all on 2e Wyckoff positions with y=0.25 and x and z for Pb (0.724, 0.581), Ti (0.761, 0.096), O1 (0.102, 0.680), O2 (0.476, 0.846), and O3 (0.144, 0.110). This is quite a dense phase, 7\% denser than $R3c$ at 80 GPa. Other structures like Cmcm post-perovskite were not stable to at least 150 GPa (Table \ref{Table:structures}). The new phase is much denser than Cmcm post-perovskite \cite{Murikami2004}, and significantly different from perovskite structures (Fig.~\ref{fig:PTO_structures}, Table \ref{Table:structures}). The Ti is 8-coordinated by O, Pb-O distances are 12\% closer than in the perovskite phases, and second neighbors Pb-Pb and Pb-Ti are closer than in the perovskite phases. Notably, this structure is not listed in the ICSD \cite{Zagorac:in5024}. It would be interesting if it shows up in other systems.

Earlier computations and experiments revealed a series of morphotropic-like phase transition at 9-22 GPa. The computations of Wu and Cohen \cite{Wu2005} considered only zone-center instabilities (five atom unit cells) and used the local density approximation (LDA) compared with GGA method in later work. It is unclear which method would be more accurate for the very small energy differences between these phases. In any event, we could be tempted to discount these earlier computations since they neglected zone-boundary instabilities, except that experiments agreed well with the predictions \cite{Ahart2008}. The fact that the room-temperature experiments reported in Ref.~\cite{Janolin2008} showed no evidence for these transitions agrees with the findings of Ref.~\cite{Ahart2008} as the latter found the transitions to occur only at low temperature ($i.e.$, 10 K). This is also consistent with the very low energy scale (less than 0.1 meV/atom) reported in Ref.~\cite{Wu2005}. Hysteresis may prevent formation of the the zone boundary $R3c$ phase, so that  the morphotropic transition is found at low temperatures, even if the zone boundary phase becomes more stable at about 10 GPa, as predicted here.

In summary, We have reexamined high-pressure ferroelectricity in PbTiO$_3$. We find consistency between theory and experiment, specifically with SHG experiments that show ferroelectricity disappearing under pressure. At high pressures a new centrosymmetric phase with $P2_1/m$ symmetry is predicted by DFT computations, which has been missed in previuos high-pressure studies of the prototype ferroelectric PbTiO$_3$. Remarkably, we find no evidence for ultrahigh pressure ferroelectricity in either experiments or theory up to megabar pressures. 

\begin{acknowledgments}
This work is supported by the U. S. Office of Naval Research (grant no. N00014-17-1-2768 and N00014-20-1-2699), U.S. Department of Energy-National Nuclear Security Administration (DOE-NNSA) through the Chicago/DOE Alliance Center (CDAC, cooperative agreement no. DE-NA0003975), and the Carnegie Institution for Science. This work was supported in part by high-performance computer time and resources from the DoD High Performance Computing Modernization Program. REC gratefully acknowledges the Gauss Centre for Supercomputing e.V. (www.gausscentre.eu) for funding this project by providing computing time on the GCS Supercomputer SuperMUC-NG at Leibniz Supercomputing Centre (LRZ, www.lrz.de). Portions of this work were performed at HPCAT (Sector 16), Advanced Photon Source (APS), Argonne National Laboratory (ANL). HPCAT operations are supported by the DOE-NNSA Office of Experimental Sciences. The APS is a DOE Office of Science User Facility operated for the DOE Office of Science by ANL (contract no. DE-AC02-06CH11357).
\end{acknowledgments}

\onecolumngrid

\begin{table}
    \centering
    \begin{tabular}{||c|c|c|c|c|c|c|c|c||}
    \hline
 & Pb-Pb & Ti-Ti & Pb-Ti & Ti-O & Pb-O & Enthalpy (eV/atom) & dH meV/atom & V \AA$^3$/atom \\
$Pm\overline{3}m$ & 6 3.61 & 6 3.61 & 8 3.13 & 6 1.81 & 12 2.55 & -877.5015 & 0 & 9.435 \\
$P4mm$ & 6 3.61 (3.60) & 6 3.61 (3.60) & 8 3.13 (3.10) & 6 1.81 (1.75) & 12 2.55 (2.47) & -877.5029 & -1.4 & 9.416 \\
$R\overline{3}c$  & 6 3.61 & 6 3.61 & 8 3.13 (3.08) & 6 1.82 & 12 2.56 (2.36) & -877.5102 & -8.6 & 9.412 \\
$R3c$ & 6 3.61 & 6 3.61 & 8 3.12 (3.02) & 6 1.82 (1.79) &  12 2.55 (2.36) & -877.5109 & -9.4 & 9.396  \\
$I4/mcm$ & 6 3.61 (3.58) & 6 3.61 (3.58) & 8 3.13 & 6 1.82 (1.81) & 12 2.56 (2.37) & -877.5113 & -9.8 & 9.406 \\
$P2_1/m$ & 4 2.78 (2.72) & 2 2.72 & 4 2.96 (2.90) & 8 1.98 (1.95) &  8 2.44 (2.08) & -877.5407 & -39.2 & 8.759 \\
$Cmcm$ (ppv) & 2 2.77 & 2 2.77 & 4 2.97 & 6 1.86 (1.85) & 9 2.39 (2.30) & -877.4180 & 83.5 & 9.281 \\  
\hline
  \end{tabular}
    \caption{Computed coordination number, average distance (\AA) (minimum distance), enthalpies and volume per atom at 80 GPa.}
    \label{Table:structures}
    
\end{table}

\newpage
\twocolumngrid

\bibliography{UHP_PbTiO3}
\end{document}